# A user-friendly nano-CT image alignment and 3D reconstruction platform based on LabVIEW *

WANG Sheng-Hao(王圣浩)[1], ZHANG Kai(张凯)[2;1)], WANG Zhi-Li(王志立)[1], GAO Kun(高昆)[1], WU Zhao(吴朝)[1], ZHU Pei-Ping(朱佩平)[2], WU Zi-Yu(吴自玉)[1,2;2)]

[1] National Synchrotron Radiation Laboratory, University of Science and Technology of China, Hefei 230027, China
[2] Institute of High Energy Physics, Chinese Academy of Sciences, Beijing 100049, China

**Abstract:** X-ray computed tomography at the nanometer scale (nano-CT) offers a wide range of applications in scientific and industrial areas. Here we describe a reliable, user-friendly and fast software package based on LabVIEW that may allow to perform all procedures after the acquisition of raw projection images in order to obtain the inner structure of the investigated sample. A suitable image alignment process to address misalignment problems among image series due to mechanical manufacturing errors, thermal expansion and other external factors has been considered together with a novel fast parallel beam 3D reconstruction procedure, developed *ad hoc* to perform the tomographic reconstruction. Remarkably improved reconstruction results obtained at the Beijing Synchrotron Radiation Facility after the image calibration confirmed the fundamental role of this image alignment procedure that minimizes unwanted blurs and additional streaking artifacts always present in reconstructed slices. Moreover, this nano-CT image alignment and its associated 3D reconstruction procedure fully based on LabVIEW routines, significantly reduce the data post-processing cycle, thus making faster and easier the activity of the users during experimental runs.

## 1. Introduction

As a unique non-destructive high resolution visualization tool, X-ray computed tomography at nanometer resolution (nano-CT) technique can be successfully applied in biomedical imaging, materials science, inspection of integrated circuits and many other scientific researches and industrial processes [1-3].

A full-field transmission X-ray microscope (TXM) with nanometer spatial resolution, named nano-CT, has been designed and assembled at the Beijing Synchrotron Radiation Facility (BSRF), a first-generation synchrotron radiation facility operating at 2.5 GeV. It operates continuously from 5 to 12 keV with fluorescence mapping capability and an achieved spatial resolution better than 30 nm [4]. In practical experiments with this nano-CT equipment, the sample stage rotates discontinuously for well defined angular intervals to obtain projection images at different angles, because it takes about 30 seconds for the CCD camera to capture a single projection image and 1 second is needed to transmit the data to the computer. During the rotation, errors in the mechanical manufacture and in the assembly would cause jittering of the rotation axis of the sample stage.

* Supported by the Major State Basic Research Development Program (2012CB825800), the Science Fund for Creative Research Groups (11321503), the Knowledge Innovation Program of the Chinese Academy of Sciences (KJCX2-YW-N42), the National Natural Science Foundation of China (NSFC 11179004, 10979055, 11205189, 11205157) and the Fundamental Research Funds for the Central Universities (WK2310000021).

1) E-mail: zhangk@ihep.ac.cn
2) E-mail: wuzy@ustc.edu.cn

Also thermal expansion due to temperature variation during an experiment and external environmental changes may impose a slight influence on the system. When dealing with micro-CT and other larger scale CT equipment, thanks to high precision micro-mechanics, these contributions are negligible. However, for the spatial resolution demands of nano-CT applications, these contributions have to be taken into account seriously, as they provoke misalignments among the projection image series, i.e. the transversal shift of the axis, axial vibrations of the sample stage, and skew phenomena. If we directly perform 3D reconstruction without an accurate calibration upon the raw dataset, all the misalignment contributions may generate blurring and streaking artifacts which cause information loss and, probably, faked structures in the reconstructed slices of the sample.

LabVIEW is a graphical programming language with a powerful function library and an easy-to-use graphic user interface (GUI) design framework. The availability of drivers for a very large number of hardware components, high-efficiency debugging functions and many other remarkable features make it an ideal software development framework for instrument-oriented programming. LabVIEW has been already used in X-ray imaging for large scale CT tomographic reconstruction software development [5], X-ray flat-panel detector driver design [6] and CT imaging system software building [7-9] by different teams. In addition, the LabVIEW's Vision Development Module (VDM) provides an excellent tool to perform digital image processing [10,11], the utilization of the VDM reduces the complexity and shortens the development cycle. The newly released LabVIEW Biomedical Workbench Toolkit (BWT) includes also a 3D image reconstruction module, which could reconstruct a 3D model from a set of 2D image slices. It provides users with a fast and reliable visualization technique for preliminary analysis and qualitative evaluations. The technique is widely used in many areas from medicine to physics and engineering, e.g., for MRI slices reconstruction. However, the BWT module cannot perform a 3D reconstruction from projection images, which prevents its use in X-ray tomographic reconstructions.

Next we will describe in detail a LabVIEW based user-friendly package designed for a high-efficiency nano-CT image alignment and a flexible 3D tomographic reconstruction.

## 2. Meterials and methods

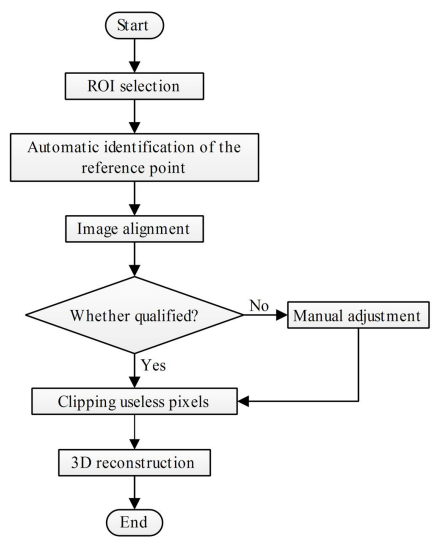

Fig. 1. The flow chart of the nano-CT image alignment and 3D reconstruction platform.

In order to fix the aforementioned image misalignment contributions affecting performance of this nano-CT equipment, a spherical gold particle with a diameter of 0.5 um is mounted on the sample stage. It is used as the reference point for the image alignment. Because of its strong absorption in the 5-12 keV range (i.e., x-ray with 12 keV photon energy has an absorption of 83.1 % through a 0.5 um thick gold plate), the projection image of the gold particle on the detector is a striking round dark area. Using this reference point present in each projection image, the software may fix the major misalignments due to the drift of the rotation axis in the plane parallel to the surface of the detector and of the sample stage's undulate position in the axial direction, while the rotation axis skew and other minor misalignments are considered negligible in our layout. The program flow chart of the nano-CT image calibration and 3D reconstruction is outlined in Fig. 1.

2.1 Selection of the region of interest

Quickly browsing the projection image series with LabVIEW using the image display indicator, it can be found that the gold particle jitters constantly within a small region. At this stage users need to choose a region of interest (ROI) including the small region containing the gold particle, where the latter image process takes place. The ROI will allow a faster and more accurate automatic location of the reference point in different projection images. A convenient and precision selection of the ROI proceeds as follows. At first, binary image corresponding to each raw projection is generated in LabVIEW with the following scheme.

If $A_{ij}=A_{min}$, set $A_{ij}=0$.

If $A_{ij}>A_{min}$, set $A_{ij}=1$.

where $A_{ij}$ is gray value of the pixel $(i, j)$ in the handled image and $A_{min}$ is the minimum gray value of the entire image. Here gray value of the pixels at the center of the reference point is assumed to be the minimum of the whole image. Actually, compared with the other region of the image, this assumption is reasonable because the central part of the gold particle strongly absorbs in the x-ray range from 5 to 12 keV. In the second step, the binary arrays of all the projection image are multiplied one by one to generate an array product, where gray values of the pixels of the gold particle's jittering area would be one, while in other parts of the image they are often zero. The method offers the possibility to easily choose an accurate ROI covering jittering area related to the gold particle.

2.2 Automatic identification of the reference point

In the choice of the ROI, two automatic methods are used to locate the reference point in each raw projection image. In the first one we position the reference point by calculating the gray value barycenter (GVB) of the subdomain after a threshold segmentation [12]. Initially, a sub image related to each projection image is extracted associated to the chosen ROI and then, by an iterative algorithm for threshold segmentation, the threshold T of each sub image is obtained and used in the next image processing step described as follows.

If $A_{ij} \leq T$, set $A_{ij} = 255 - A_{ij}$.

If $A_{ij} > T$, set $A_{ij} = 0$.

Assuming that the raw projection images are 8-bit unsigned and the procedure is valid for each projection image, the GVB can be calculated according to Eq. 1. This set of coordinates is considered to be the center of the reference point in the selected projection image.

$$\begin{cases} X=\bar{i} = \dfrac{\sum A_{ij} \times i}{\sum A_{ij}} \\ Y=\bar{j} = \dfrac{\sum A_{ij} \times j}{\sum A_{ij}} \end{cases} \quad (1)$$

The second automatic identification procedure of the reference point is the circle fitting method (CFM) based on the image morphology. It fully takes advantage of the LabVIEW's VDM module, a customized shape detection function is used to identify a geometrical shape in an image, and to acquire the position of the reference point.

### 2.3 Image alignment

Based on the automatic procedure for the acquisition of the coordinates of the reference point in each projection image, the image alignment process is carried out as follows. In the horizontal direction, the symmetry axis of the image is considered the alignment target of the reference point in all projection images, while in the vertical direction, the reference point in the first projection image is regarded as the alignment target of all others. And then, via a small shift of each projection image in both directions and with the aim of aligning the reference point with the target, the image calibration procedure would be completed. However, if we still perform the image alignment in the horizontal direction with the above mentioned strategy, when the gold particle is mounted slightly away from the rotation axis of the sample stage, in the horizontal direction a very narrow view is generated due to a large scale pixels movement. As a consequence we consider an alternate method in which the pixel position X(k) in the horizontal direction, described in Eq. 2 is set as the alignment target of the reference point in the $k^{th}$ projection image:

$$X(k) = \frac{D}{2} - R \times \cos\frac{(k-1)\pi}{N}. \tag{2}$$

Here D is the pixel width of the raw projection image and R is the pixel distance between the reference point and the horizontal symmetry axis in the first projection image. The image calibration in the horizontal direction can be performed in this way without a large shift of the pixels.

### 2.4 Manual adjustment module

Because the characteristics of the reference point may change significantly among different experiments, and in some circumstances neither GVB nor CFM generate a satisfactory alignment upon the raw projection image series, although time-consuming, a manual correction module (MCM) with a sub-pixel positioning accuracy has been introduced. It represents a relevant support to the automatic protocols.

Finally, after image alignment of the raw nano-CT dataset, useless pixels in each projection image resulting from the image shift are clipped, both to decrease the consumed memory and to minimize the time required in the following 3D reconstruction.

### 2.5 3D reconstruction

The 3D tomographic reconstruction is performed here by the inverse radon transform function *iradon* of the MATLAB software. LabVIEW's MATLAB script node is used to invoke the MATLAB software script server to execute scripts written in the MATLAB language syntax, and the reconstructed slice is displayed with three basic orthographic views.

### 3. Results

Fig. 2 is the main GUI of the nano-CT image alignment and the 3D reconstruction software platform. It is made up of a 4 pages tab control representing the four key procedures, i.e. ROI selection, automatic correction, manual correction and 3D reconstruction, respectively. Fig. 2(a) shows the GUI of the ROI selection, after the program's execution, the projection images series of the raw CT dataset is shown successively in the left image display indicator. During the process users may quickly check the correction of the raw dataset and may also evaluate qualitatively the jittering area of the gold particle. Then the *cumprod* (cumulative production) image will appear on the right image display indicator, where reference points in all raw projection images form a dark trail. A brown rectangle covering the dark trail could be drawn as showed in Fig. 2(a), then each projection image with the over imposed rectangle, would be shown in the left image display indicator, allowing user to check the accuracy of the selected ROI and perform further adjustments.

The GUI of the automatic correction procedure is showed in Fig. 2(b). Users may choose one automatic identification method between GVB and CFM. When dealing with each projection image, a white circle with a radius matching the gold particle would overlay at the automatically identified position, to assess the automatic identification performance. The radius of the white circle can be customized to optimize the contrast respect to the reference point, while in some cases a multiple threshold segmentation is required.

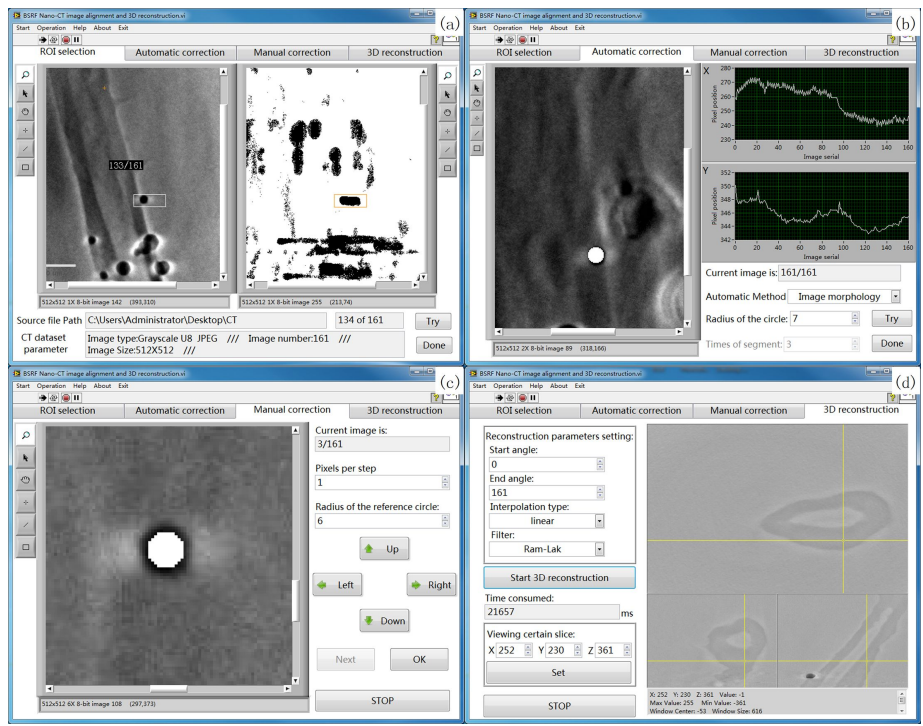

Fig. 2. (color online) GUI of the nano-CT image alignment and 3D reconstruction platform. Panels (a) (b) (c) (d) show the procedures of ROI selection, automatic correction, manual correction and 3D reconstruction, respectively.

Fig. 2(c) represents the GUI of the manual calibration module introduced as an important complementary tool of the automatic procedure. Actually, an image can be manually shifted in four directions: up, down, left and right by a certain amount of pixels, and shortcut keys corresponding to the four directions of shift are enabled. A white circle, appearing in the image at the expected position, acts as the alignment target of the reference point. A sub-pixel positioning accuracy can be achieved by an optimal matching between the reference point and the target as shown in Fig. 2(c). The radius of the circle can be customized to achieve a better contrast and also pixels per step can be changed to make possible and easier the adjustment when the reference point is located far from the target.

Fig. 2(d) shows the GUI of the 3D reconstructor. In the left part, reconstruction parameters such as start-stop projective angle, interpolation type (*nearest*, *linear*, *spline*, *pchip*, *cubic*, *v5cubic*) and filter (*Ram-Lak*, *Shepp-Logan*, *Cosine*, *Hamming*, *Hann*, *None*) can be customized. After the 3D tomographic reconstruction, reconstructed slices appear in the three basic orthographic views, while other cross-sections of the inspected material can be viewed by dragging the vernier in the image display indicator or can be set to a definite three-dimensional coordinate.

## 4. Performance

### 4.1 Image alignment efficiency

To evaluate the performance of the nano-CT image alignment package and the 3D tomographic reconstructor, we present here some experimental results obtained at BSRF. The first nano-CT raw dataset consists of 181 projections of 8 bit unsigned 1024 1024 images, and the inspected sample is a standard spherical gold particle with a diameter of 4.5 um. During the image alignment process, GVB and MCM image alignment methods are independently applied to correct the nano-CT projection image series. Finally, the *linear* interpolation and *Ram-Lak* filter are utilized for tomographic reconstruction. Fig. 3(a) shows the slice reconstructed from raw projection images, while Fig. 3(b) and Fig. 3(c) show the slices reconstructed from corrected datasets after GVB and MCM image calibrations, respectively. It is clear that a serious blurring and evident additional fake artifacts exist in the slice reconstructed from the raw dataset as shown in Fig. 3(a), which have been remarkably improved in the reconstruction image after an image alignment procedure. And we can also see that almost no obvious visual discrepancies occur between the slices reconstructed from projection images calibrated with GVB and MCM strategy.

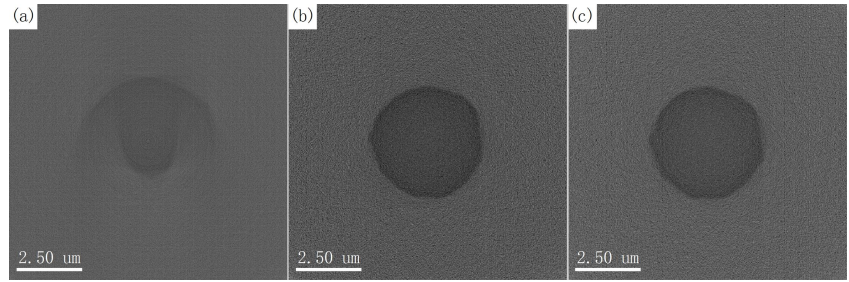

Fig. 3. A cross section of the gold particle reconstructed with (a) the raw projection images, and projection image series corrected independently with (b) the GVB and (c) the MCM calibration procedures.

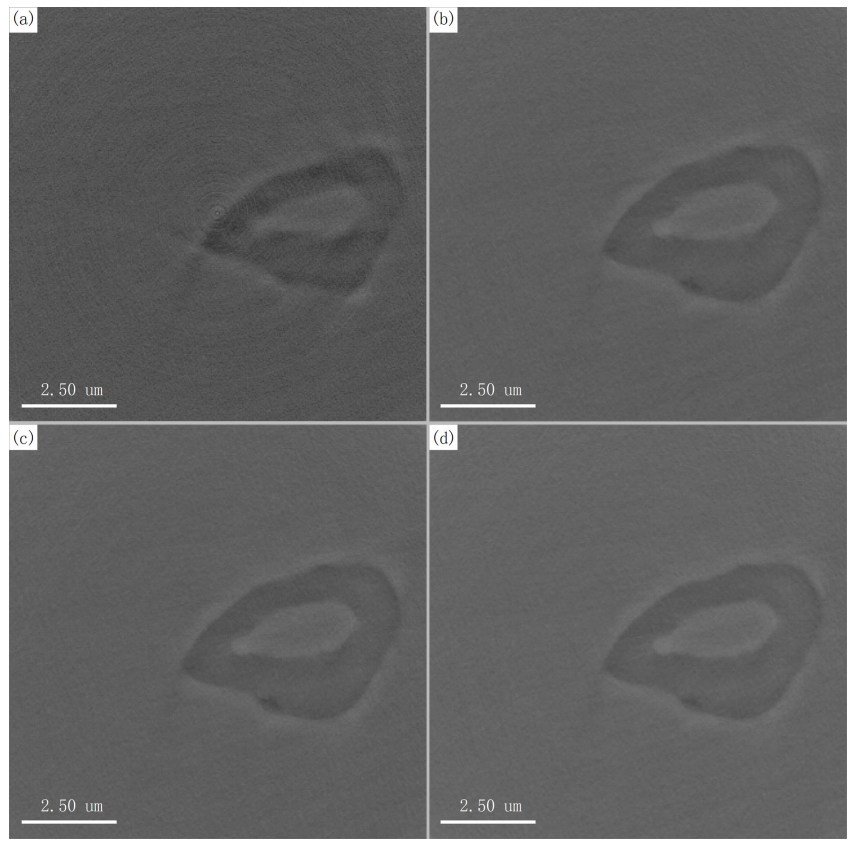

Fig. 4. A transverse slice of an animal ovary, data are reconstructed from (a) raw projection images, and projection images corrected independently with (b) the GVB, (c) the CFM and (d) the MCM calibration procedures.

The second dataset is made up of 161 projections of 8-bit unsigned 512 512 images and the material under investigation is the ovary of an insect. During the nano-CT image alignment procedure, GVB, CFM and MCM image calibration strategy procedures are used independently. Moreover, a *linear* interpolation and a *Ram-Lak* filter are utilized for a tomographic reconstruction. The reconstructed transverse slices are shown in Fig. 4, while in Fig. 5 a coronal slice through the vagina is showed. It is very clear as shown in Fig. 4(a) and Fig. 5(a) that evident ring artifacts and blurring occur in the slices reconstructed from raw projection images, while almost ideal reconstruction results are obtained after the image calibration with one of the three available strategies. A comparison of the quality of the reconstructed slices from data corrected with GVB, CFM and MCM points out that there are no obvious differences between them, except for the existing slight blurring in the boundary of the coronal slice through the vagina (see Fig. 5(b)).

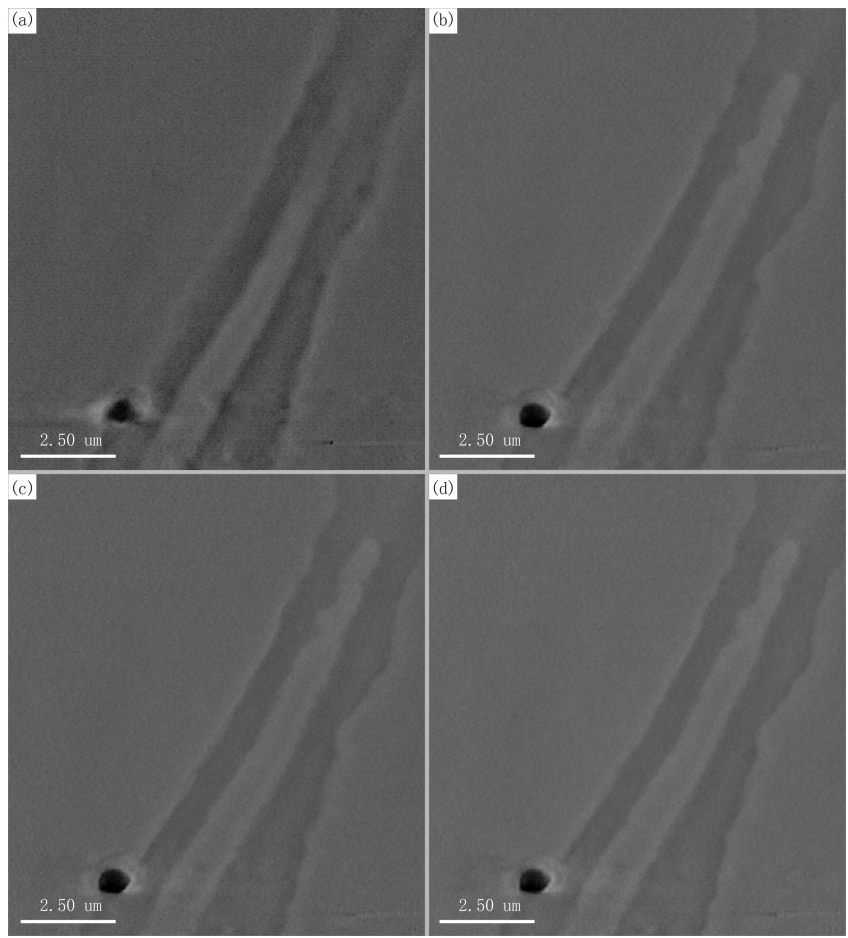

Fig. 5. A coronal slice through the vagina of an insect, data are reconstructed from (a) raw projection images, and projection images corrected independently with (b) the GVB, (c) the CFM and (d) the MCM calibration procedures.

### 4.2 Time consumed in image alignment and 3D reconstruction

Table 1. Time required for nano-CT image alignment and 3D reconstruction of two datasets

| Dataset | | | Image alignment with an automatic /manual procedure (s) | 3D reconstruction (s) |
|---|---|---|---|---|
| 181 | 1024 | 1024 | 166.6/851.3 | 136.2 |
| 161 | 512 | 512 | 106.3/805.8 | 19.6 |

Reported in Table 1 are the values of the time consumed to perform the nano-CT image alignment and the 3D reconstruction of the two raw datasets we discussed above. The evaluation has been performed on a Core i7 3.40 GHz personal computer with 16 GB RAM and 64-bit windows 7 operating system. The reconstruction results of these two datasets are 1024 slices of 724 724 pixels and 512 slices of 362 362 pixels, respectively. It takes 19.6 seconds to reconstruct a 512 362 362 array, which could be considered a real breakthrough in increasing the reconstruction speed, excluding other performance factors, compared with the existing LabVIEW-based 3D Reconstructor [5].

## 5. Discussion

Two automatic correction algorithms/methods and a manual adjustment module are described in this manuscript to perform the nano-CT image calibration before a tomographic reconstruction. Both automatic methods have advantages and drawbacks in different conditions, GVB is faster, has less possibility to miss the target, and its positioning deviation is about 3-4 pixels. On the contrary, CFM is slower, but it may identify more accurately the position, especially when the reference point is not symmetric. However, CFM more frequently misses the reference point, a condition that needs a supplementary manual adjustment. By combining GVB and CFM, the system is much more powerful, in particular, in complex situations. The sub-pixel positioning accuracy of the MCM calibration strategy makes it the most reliable image correction method if not for time consideration, because about 15 minutes are necessary to handle a set of CT raw dataset containing 181 projections. Fortunately, as shown in our experimental results, almost no observable differences are detected in the slices reconstructed from datasets with an automatic correction or a manual calibration method. Typically, about 2-4 pixels positioning deviation of the reference point would exist after the two automatic alignment procedures, which normally is considerately tolerant for the latter 3D tomographic reconstruction.

The automatic image alignment strategy we described allows labor-saving and fast data pre-processing before the tomographic reconstruction. while the sub-pixel positioning accuracy but time-consuming MCM module is supposed to only act as a complementary tool in challenging situations where neither GVB nor CFM could effectively identify the reference point.

Several reasons are behind the choice of LabVIEW as the program development environment. The first one is that LabVIEW is a graphical rather than a text based programming language such as $C^{++}$ or FORTRAN, which allow a much faster and easier software development. The second one is that LabVIEW's VDM module provides users with a unique and ideal digital image processing tool, which remarkably reduces the project development cycle. Finally, LabVIEW integrates perfectly with MATLAB, a choice that allows developers to fully take advantage of the MATLAB's comprehensive library function and its superior computation ability. In our case, by constructively invoking the inverse radon transform function *iradon* of MATLAB with the LabVIEW's MATLAB script node, complex codes of parallel beam tomographic reconstruction are avoided. Moreover, the convenient multithreaded programming technique in LabVIEW makes developers capable to fully exploit the multiple Central Processing Units of their computers.

## 6. Conclusion

A LabVIEW based reliable and user-friendly nano-CT image alignment and 3D reconstruction software package is presented and discussed. We have showed how instability problems associated to rotation axis of the sample stage can be overcome, minimizing artifacts in the 3D reconstruction. The nano-CT image alignment and 3D reconstruction software package optimized for the nano-CT equipment existing at BSRF, significantly shortens the data post-processing cycle, thus making easier the users activity during runs, and it also has great potentials for the upgrade of other existing nano-CT instruments.

**Acknowledgements**

The authors acknowledge A. Marcelli for many fruitful helps and discussions.